\documentclass[useAMS,fleqn,usenatbib]{mnras}

\usepackage[T1]{fontenc}
\usepackage{ae,aecompl}

\usepackage{amsmath}
\usepackage{amssymb}
\usepackage{graphicx}
\usepackage{natbib}
\usepackage[version=3]{mhchem}
\usepackage{setspace}
\usepackage{float}
\usepackage{mathrsfs}
\usepackage{subfigure}
\title[AGB stars in NGC\,6397]{AGB subpopulations in the nearby globular cluster NGC\,6397}

\author[B. T. MacLean, et al.]{B. T. MacLean$^{1}$\thanks{E-mail: ben.maclean@monash.edu}, 
S. W. Campbell$^{1,2}$, 
G. M. De Silva$^{3,4}$, 
J. Lattanzio$^{1}$, 
\newauthor V. D'Orazi$^{7,5,6}$, 
P. L. Cottrell$^{1,8}$, 
Y. Momany$^{7}$ and 
L. Casagrande$^{9}$ \vspace{2mm}\\
$^{1}$Monash Centre for Astrophysics, School of Physics and Astronomy, Monash University, Victoria 3800, Australia\\ 
$^{2}$Max-Planck-Institut f{\"u}r Astrophysik (MPA), Karl-Schwarzschild-Strasse 1, 85748 Garching, Germany \\ 
$^{3}$Australian Astronomical Observatory, 105 Delhi Rd, North Ryde, NSW 2113, Australia \\
$^{4}$Sydney Institute for Astronomy, School of Physics, The University of Sydney, NSW 2006, Australia \\
$^{5}$Research Centre for Astronomy, Astrophysics \& Astrophotonics (MQAAAstro), Macquarie University, Sydney, NSW 2109, Australia \\ 
$^{6}$Department of Physics and Astronomy, Macquarie University, North Ryde, NSW 2109, Australia \\
$^{7}$INAF Osservatorio Astronomico di Padova, vicolo dell'Osservatorio 5, I-35122, Padova, Italy \\
$^{8}$School of Physical and Chemical Sciences, University of Canterbury, Private Bag 4800, Christchurch 8140, New Zealand \\
$^{9}$Research School of Astronomy and Astrophysics, Australian National University, Canberra, ACT 2611, Australia
}
\date{Accepted TBC. Received TBC; in original form TBC}

\begin{document}

\pagerange{\pageref{firstpage}--\pageref{lastpage}} \pubyear{\the\year}

\maketitle

\label{firstpage}

\begin{abstract}
It has been well established that Galactic Globular clusters (GCs) harbour more than one stellar population, distinguishable by the anti-correlations of light element abundances (C-N, Na-O, and Mg-Al). These studies have been extended recently to the asymptotic giant branch (AGB). Here we investigate the AGB of NGC\,6397 for the first time. We have performed an abundance analysis of high-resolution spectra of 47 RGB and 8 AGB stars, deriving Fe, Na, O, Mg and Al abundances. We find that NGC\,6397 shows no evidence of a deficit in Na-rich AGB stars, as reported for some other GCs -- the subpopulation ratios of the AGB and RGB in NGC\,6397 are identical, within uncertainties. This agrees with expectations from stellar theory. This GC acts as a control for our earlier work on the AGB of M\,4 (with contrasting results), since the same tools and methods were used.
\end{abstract}



\section{Introduction}

It is well known that Galactic globular clusters (GCs) show star-to-star spreads in the abundances of proton-capture elements (primarily He, C, N, O and Na), while most GCs remain homogeneous in the iron peak species \citep{carretta2009}. This spread often presents as multi-modal \citep[as in the early low-resolution cyanogen (CN) studies of][]{norris1981m4,cottrell1981}, with two or more distinct subpopulations being identified. One of these subpopulations is always chemically similar to Galactic halo stars of the same metallicity -- designated here as SP1 and which is inferred to contain primordial He abundances -- with one or more further subpopulations found to have higher N and Na (and lower C and O) abundances -- here designated collectively as SP2 \citep[see][for an extensive review]{gratton2012}. These are the ubiquitous C-N and Na-O (and Mg-Al in some GCs) anti-correlations \citep{carretta2009vii}. This spread in light elemental abundance can also be inferred from narrow and intermediate band photometric data, seen as multiple red- or sub-giant branches, or multiple main sequences in a GC's colour-magnitude diagram \citep[e.g.,][]{milone2008,milone2014}.

The peculiar abundance signature of SP2 stars has been observed in both evolved and unevolved stars in many clusters \citep{gratton2001}, indicating that this pattern is likely to have been inherited at birth. Furthermore, the pattern is generally not observed elsewhere, such as the (less massive) open clusters of the Galaxy \citep{desilva2009,maclean2015}; however very recently it has been suggested that the Galactic bulge may contain SP2-like stars {\mbox{\citep{schiavon2017}}}. The most common explanation for this light-elemental inhomogeneity is the self-pollution hypothesis where the ejecta of more massive SP1 stars mixed with an early dense interstellar medium, from which SP2 stars were formed \citep{cannon1998,gratton2004}.

Importantly, the relative fractions of each subpopulation remain the same through all these phases of evolution, as expected from stellar evolutionary theory. However, until recently there were no systematic surveys of asymptotic giant branch (AGB) stars. Some early \citep[e.g.,][]{norris19816752} and more recent \citep{campbell2010,campbell2013} low-resolution spectroscopic studies of GCs found that the distribution of cyanogen band strengths varies greatly between the RGB and AGB of several GCs. In particular, they found no CN-strong (i.e., SP2) AGB stars in NGC\,6752, which has an extended blue horizontal branch (HB). These results hinted at differences in evolution between stars of different light elemental abundances, which are not fully predicted in standard stellar evolution theory -- only stars with extreme He abundances are expected to avoid the AGB phase due to smaller envelopes in the HB phase \citep{dorman1993,campbell2013,cassisi2014}.

In this paper we use the prescription as described in \citet[hereafter ML16]{maclean2016}, where the percentages of RGB and AGB stars in a GC that are found to be members of SP2 are written as $\mathscr{R}_{RGB}$ and $\mathscr{R}_{AGB}$, respectively \citep[typical $\mathscr{R}_{RGB}$ values are {$\sim$}50-70\%;][]{carretta2010}; and the SP2 AGB deficit is given by
\begin{equation} \mathscr{F} = (1-{\frac{\mathscr{R}_{\rm AGB}}{\mathscr{R}_{\rm RGB}}}){\cdot}100\%, \end{equation}
where a value of 100\% indicates that no SP2 stars reach the AGB -- as reported for NGC\,6752 and M\,62 by \citet{campbell2013} and \citet{lapenna2015}, respectively. For clusters with extended HBs (where the bluest stars reach $T_{\rm eff}$ over 15,000~K; e.g., NGC\,6752, NGC\,2808), an $\mathscr{F}$ value of up to $\sim$30\% may be expected due to the well-established existence of AGB-manqu\'{e} stars \citep[which evolve directly from the HB to the white dwarf phase, avoiding the AGB;][]{agb-manque,dorman1993,cassisi2014}. Clusters whose HBs do not extend into this regime (e.g., M\,4, NGC\,6397) are expected to have an $\mathscr{F}$ value of zero per cent, with all stars in the cluster ascending the AGB.

There has been much debate as to the level and existence of GC SP2 AGB deficits in recent years as more evidence has been gathered, but a definitive conclusion has yet to be reached. In fact, contradictory evidence has been presented for both NGC\,6752 and M\,4. For example, in \citet{campbell2013} we found that the measured Na abundances in all NGC\,6752 AGB stars were consistent with SP1, indicating $\mathscr{F} \sim 100\%$. \citet{lapenna2016} conducted an independent study of the same GC, and found that with [Na\,{\sc i}/Fe\,{\sc i}] abundances, $\mathscr{F}$ dropped to the predicted value of $\sim$30\%. The assumption that dividing by Fe\,{\sc i} is more accurate has recently been disputed by \citep[hereafter C17]{campbell2017}. Recent studies of AGB stars in other GCs such as \citet{johnson2015}, \citet{garciahernandez2015}, and \citet{wang2016} have found varying values of $\mathscr{F}$ -- see Table 4 of ML16 for a summary of $\mathscr{F}$ values as of July 2016.

Attempts to theoretically explain SP2 AGB deficits have been outpaced by the numerous observational studies that have painted a complex picture, both technically (e.g., the treatment of non-LTE) and empirically (e.g., contradictory results). If high SP2 AGB deficits are real, rather than being an artefact of the spectroscopic analysis (see \S\ref{6397_discussion} for discussion), then the most likely explanation comes from the He-enrichment of SP2 stars. This results in smaller envelope masses on the HB \citep{gratton2010,cassisi2014,charbonnel2015,charbonnel2016} and such stars are known to evolve directly to the white dwarf phase (AGB-manqu\'{e} stars). SP2 AGB deficits above $\mathscr{F} \simeq 30\%$ suggest that the location along the HB where this alternative evolutionary path begins to occur may be incorrectly predicted by theory, and/or dependent on more factors than previously thought.

Similar to the debate on AGB abundances in NGC\,6752, recent studies on the archetypal GC M\,4 have presented starkly different conclusions on the nature of its AGB. ML16 presented [Na/Fe] and [O/Fe] abundances for both AGB and RGB stars in M\,4, reaching the conclusion that all AGB stars are consistent with being SP1 stars (i.e., $\mathscr{F} \simeq 100\%$). In contrast, \citet{lardo2017} and \citet{marino2017} -- using photometric indices and spectroscopic analysis, respectively -- concluded that the spread of light elemental abundances in the AGB of M\,4 is similar to the RGB (however, both studies found that their AGB samples were offset toward SP1-like abundances). If true, this is consistent with the theoretical prediction of $\mathscr{F} = 0\%$. However, the very recent study of \citet{wang2017} showed that the spread in  Na abundances of M\,4's AGB is significantly narrower than the RGB, qualitatively similar to the findings of ML16, but not as extreme. It is clear that further study of this GC is required.

If high SP2 AGB deficits are reliably demonstrated, this may impose new and important restrictions on low-mass, low-metallicity stellar evolution and/or atmospheric models; impacting the field of globular clusters, stellar evolution, and Galactic formation and archaeology. 

In the current study we aim to derive AGB subpopulation ratios for the GC NGC\,6397 for the first time. NGC\,6397 is an old and metal-poor GC with a well documented Na-O anti-correlation on the RGB, the range of which is smaller than many other clusters \citep[no `extreme population' in the classification of][which is associated with high He abundance]{carretta2009}. NGC\,6397 also displays a Mg-Al anti-correlation \citep[hereafter L11]{lind20116397}. The short (but blue) HB of NGC\,6397 extends between $8000 K<T_{\rm eff}<10,500 K$, suggesting that no stars in the cluster should evolve into AGB-manqu\'{e} stars \citep{lovisi2012}. In order to determine if this is the case, we have performed an analysis of spectra from a sample of AGB and RGB stars in NGC\,6397. For each star we have derived radial velocities, stellar parameters, and abundances of Fe, Na, O, Mg and Al.

\section{Sample selection, observations and membership}
\label{6397_obs_member}

Our stellar targets were selected from the NGC\,6397 photometric dataset of \citet[UBVI from the ESO/MPG WFI, see Table~\ref{tab:6397_obs}]{momany2003}. For the bright stars considered here the photometric completeness is $100\%$, for all colours. The photometry covers the entire cluster out to at least 9 arcmin from the cluster centre (in some directions reaching to $\sim 22$ arcmin). This compares with the cluster's half-light radius of 2.9 arcmin \citep{harris1996}. To avoid crowding problems in the core with multi-object fibre placement the sample was limited to stars outside $\sim 0.5$ arcmin of the cluster centre.

\begin{figure}
\centering
\includegraphics[width=0.9\linewidth]{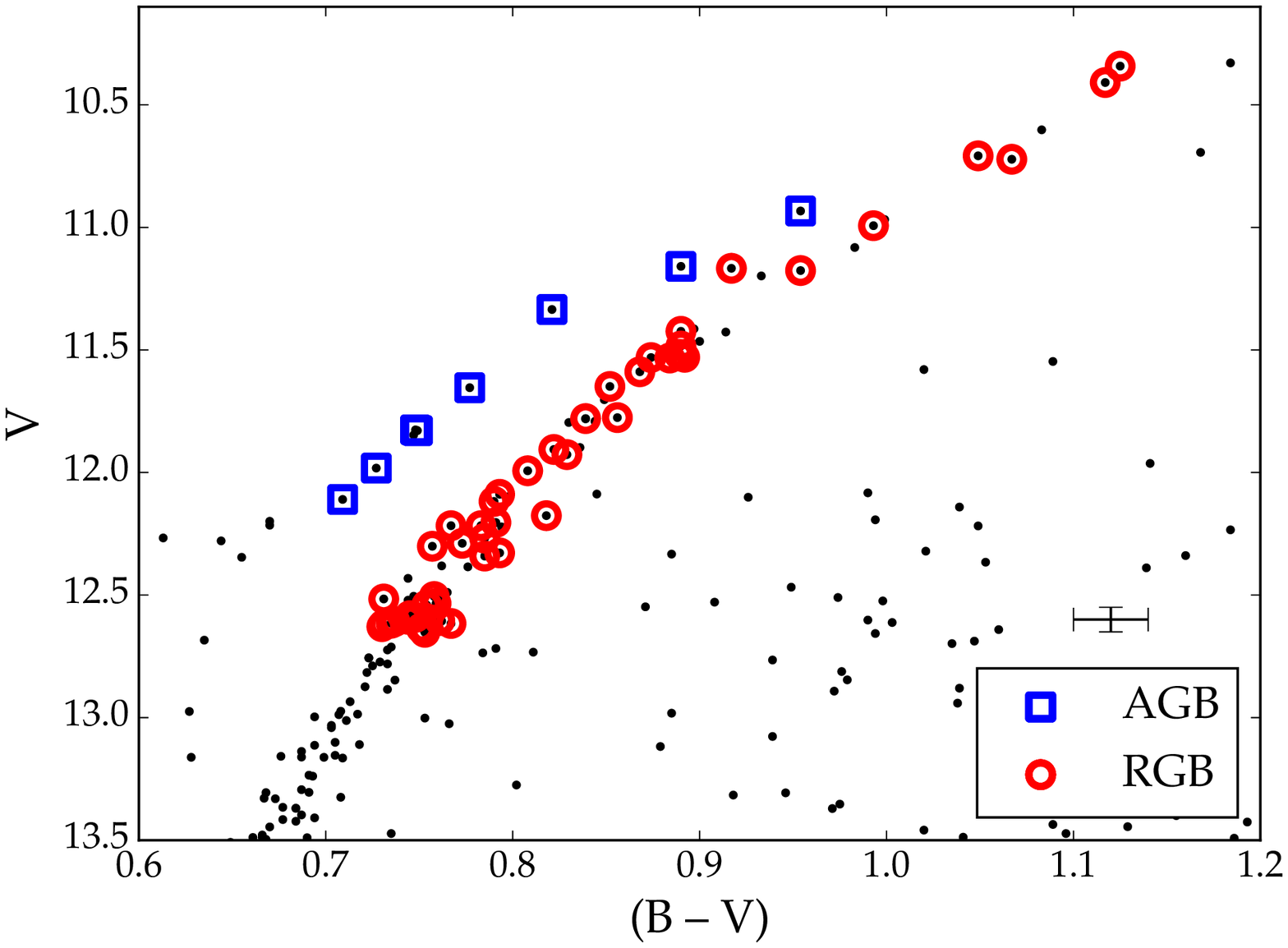}
\includegraphics[width=0.9\linewidth]{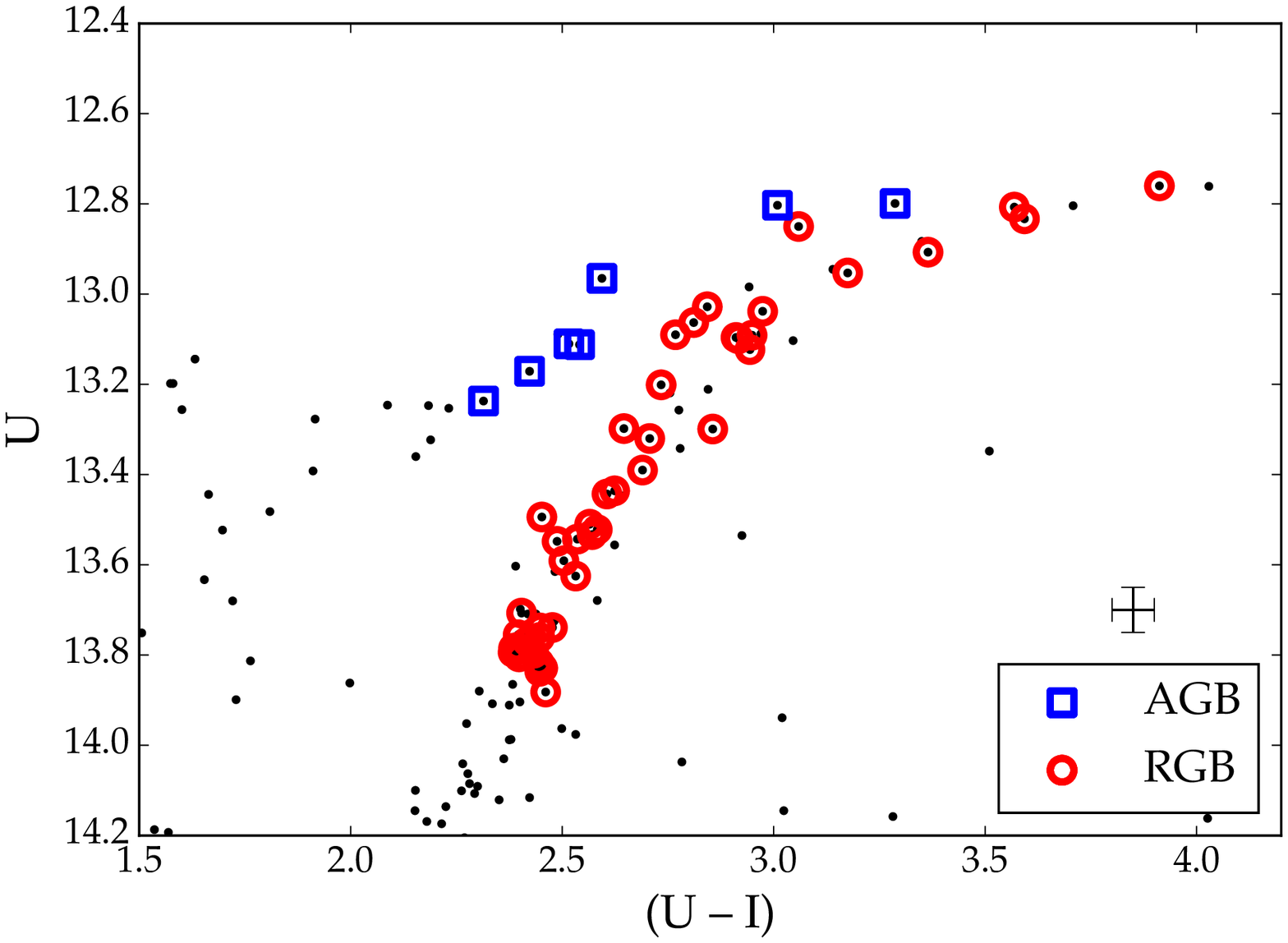}
\caption{V$-$(B$-$V) and U$-$(U$-$I) colour-magnitude diagrams of the observed NGC\,6397 RGB and AGB stars (open circles and squares, respectively), displayed over the full photometric sample of \citet[black points]{momany2003}. In the top panel, a constant reddening correction value of (B$-$V) = $-0.19$ was applied to all photometric data. No reddening correction was applied to the (U$-$I) photometry (bottom panel). We note that there are only 7 AGB stars in the U$-$(U$-$I) diagram because one star (AGB 80621) does not have a reliable U-band magnitude and was selected based only on its B- and V-band magnitudes.}
\label{fig:6397_cmds}
\end{figure}

The RGB and AGB are separated in V$-$(B$-$V) and U$-$(U$-$I) space (Figure~\ref{fig:6397_cmds}). AGB stars were conservatively selected -- only early-AGB stars were included so as to avoid the mislabelling of stars since the AGB and RGB colours become similar at brighter magnitudes. We then cross-matched our selection with the 2MASS database to take advantage of the high quality astrometry and JHK photometry. 2MASS IDs and JHK photometric magnitudes for the whole sample are included in Table~\ref{tab:6397_obs}. In total our initial target sample included 9 AGB stars and 64 RGB stars. Importantly for the science goal of this study the RGB and AGB samples are spatially coincident.

High-resolution spectra were collected in July 2015 using 2dF+HERMES on the Anglo-Australian Telescope which provides $R=28,000$ spectra in four narrow windows; blue (4715 - 4900\AA), green (5649 - 5873\AA), red (6478 - 6737\AA), and infrared (7585 - 7887\AA) \citep[for more details on the HERMES instrument, see][]{galah,sheinis2015}. Due to restrictions on 2dF fibre positioning, we were able to collect spectra for only 60 of the 73 targets. This down-sampling is random, except that priority was given to obtaining the largest possible sample of AGB stars, since the number of AGB stars is inherently low compared to RGB stars (see Fig~\ref{fig:6397_cmds}, black dots). In total we collected spectra for 8 of the 9 identified AGB stars, and 52 RGB stars. 

The spectra had an average signal-to-noise ratio of 70. The software package {\sc 2dfdr} \citep[v6.5]{2dfdr} was used to reduce the data for analysis. Radial velocities were measured with the {\sc iraf} \textit{fxcor} package \citep{tody1986}, using a solar reference template. The mean radial velocity for NGC\,6397 after non-member elimination was found to be {\textless}v{\textgreater} = $19.30 \pm 0.48$~km/s (${\sigma}=3.71$km/s), consistent with \citet{lind2009}, who report {\textless}v{\textgreater} = $18.59 \pm 0.16$~km/s (${\sigma}=3.61$~km/s). Individual stellar radial velocities are listed in Table~\ref{tab:6397_obs}. Iterative 3-$\sigma$ clipping of radial velocities and metallicities (discussed in $\S$\ref{6397_abund_method}) reduced the final RGB sample to 47 stars. All of the 8 observed AGB stars were found to be members.

Apart from not sampling the inner core of the cluster we do not identify any sample bias. Moreover we have collected spectra for almost all of the AGB stars in the very wide field of view of the source photometry. The 47 RGB stars offer a solid basis for comparison. The final observed samples can be seen visually in the colour-magnitude diagrams of Figure~\ref{fig:6397_cmds}, over-plotted against the full photometry sample.

\begin{table*}
\centering
\caption{NGC\,6397 target details including data from \protect{\citet[UBVI photometry and target IDs]{momany2003}} and 2MASS \protect{\citep[JHK photometry -- gaps in data represent targets with low quality flags]{2mass}}, radial velocities (km/s), and \protect{\citet[L11]{lind20116397}} IDs. Full table available online.}
\label{tab:6397_obs}
\begin{tabular}{cccccccccccc}
\hline
ID & Type & 2MASS ID         & L11 ID & V Mag & B Mag & U Mag & I Mag & J Mag & H Mag & K Mag & RV (km/s)\\
\hline\hline
56897 & AGB  & 17400665-5335001 & -      & 11.83  & 12.76  & 10.59  & 13.11  & 9.76   & 9.25   & 9.13   & 17.17  \\
60609 & AGB  & 17402547-5347570 & -      & 11.65  & 12.62  & 10.37  & 12.97  & -      & -      & -      & 20.68  \\
70509 & AGB  & 17405254-5341049 & -      & 11.98  & 12.90  & 10.75  & 13.17  & 9.95   & 9.48   & 9.31   & 19.38  \\
70522 & AGB  & 17404076-5341046 & -      & 11.16  & 12.24  & 9.79   & 12.80  & 8.94   & 8.37   & 8.26   & 18.93 \\
73216 & AGB  & 17403510-5339572 & -      & 11.83  & 12.76  & 10.57  & 13.11      & -   & -   & -   & 16.00  \\
\vdots & \vdots & \vdots & \vdots & \vdots & \vdots & \vdots & \vdots & \vdots & \vdots & \vdots & \vdots \\
\hline
\end{tabular}
\end{table*}

\section{Method}

\subsection{Atmospheric parameters}

For this study we have used several photometric relations to determine effective temperatures for all stars. 

Typically with spectroscopic studies (such as ML16), stellar parameters are determined by requiring the excitation and ionisation balance of abundances from neutral and singly-ionised iron (Fe\,{\sc i} \& Fe\,{\sc ii}, respectively) absorption lines \citep[e.g.,][]{sousa2014}. While a significant strength of this method is that the parameters are unaffected by photometric reddening, there are also many weaknesses. Many solutions can be found for a single star, largely depending on the choice of initial parameter estimates (see C17). Additional spectroscopic uncertainties such as EW measurements, choice of atmospheric model, atomic line data, and parameter interdependence can compound this problem.

To further complicate the picture, \citet{lapenna2014,lapenna2016} have provided evidence that the Fe\,{\sc i} lines of AGB stars may experience a higher degree of non-LTE effects than RGB stars at the same metallicity and effective temperature. If true, then assuming ionisation balance may artificially and preferentially lower the derived surface gravity of AGB stars \citep{lind2012}. In C17 we suggested that this so-called `AGB iron over-ionisation problem' does not exist (at least in NGC\,6752), but may be the result of systematic offsets in photometrically-derived $T_{\rm eff}$. Regardless, Fe\,{\sc i} lines are well known to experience some non-LTE effects \citep[on both the RGB and AGB, and especially at low metallicities, see][]{bergemann2012}, so forcing ionisation balance prior to the correction of non-LTE effects may result in systematically incorrect gravities and metallicities in all stars.

We have used the B$-$V and V$-$K relations from \citet{ramirez2005}, \citet{gonzalez2009} and \citet{casagrande2010} to determine $T_{\rm eff}$ estimates. Additionally, we have calculated $T_{\rm eff}$ without relying on colour calibrations, by implementing the infrared flux method (IRFM) at an estimated log $g$ of each AGB and RGB star, as described in \citet{casagrande2010,casagrande2014} using BVI and 2MASS JHK photometry. Thus we have seven $T_{\rm eff}$ estimates for each star. These methods are dependent on metallicity, for which a value of [Fe/H] $= -2.00$ was assumed for NGC\,6397. To account for interstellar extinction we applied a constant correction of $E(B-V)=-0.19$ to all stars \citep{gratton2003}. NGC\,6397 does not suffer from significant differential reddening \citep{milone20126397}.

\begin{figure}
\centering
\includegraphics[width=0.9\linewidth]{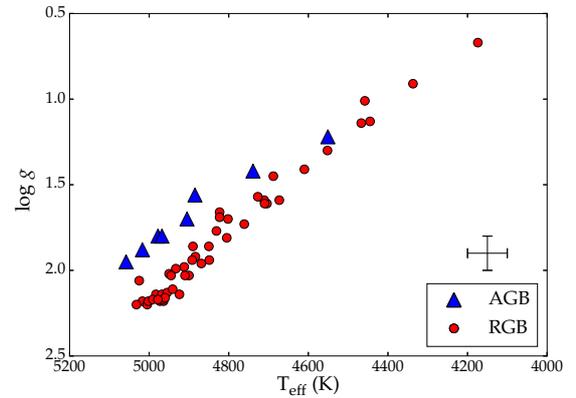}
\caption{Final stellar parameters of NGC\,6397, determined from photometric relations. The method of parameter determination is described in the text. Typical uncertainties are indicated, and are the same as in Table~\ref{tab:6397_atmos}.}
\label{fig:6397_hr}
\end{figure}

Four stars were flagged for low quality and/or contamination in the 2MASS database so only the B$-$V relations were used to determine $T_{\rm eff}$ for these stars. For all other stars, the mean of the seven $T_{\rm eff}$ estimates was adopted. Table~\ref{tab:6397_teff_diffs} shows the variation between the final adopted $T_{\rm eff}$ values and those of the photometric relations and IRFM. 
Surface gravities (log $g$) and micro-turbulences ($v_t$) were determined using the empirical relations from \citet{alonso1999} and \citet{gratton1996}, respectively, and assuming a mass of 0.8~$M_{\odot}$ and 0.7~$M_{\odot}$ for the RGB and AGB, respectively \citep{lovisi2012,miglio2016}. We adopt a 1$\sigma$ uncertainty of $\pm$50K for T$_{\rm eff}$ (see Table~\ref{tab:6397_teff_diffs}), $\pm$0.1 dex for log $g$, and $\pm$0.2 km/s for $v_t$. Final stellar parameters for each star are included in Table~\ref{tab:6397_abunds} and represented visually in Figure~\ref{fig:6397_hr}.

\begin{table}
\centering
\caption{Average differences in T$_{\rm eff}$ between the adopted value and each photometric estimate. Uncertainties are the 1$\sigma$ standard deviations of the cluster samples. The average $\sigma$ value in the last row is indicative of the spread of T$_{\rm eff}$ estimates for each star.}
\label{tab:6397_teff_diffs}
\begin{tabular}{lcc}
\hline
Method & ~~~~$\Delta$T$_{\rm eff}$ (K)  \\
\hline\hline
Ram (B$-$V)$^1$            & ~~~~$94\pm45$     \\
Gonz (B$-$V)$^2$           & ~\,$-17\pm42$  \\
Casa (B$-$V)$^3$            & ~~~~$22\pm98$ \\
Ram (V$-$K)$^1$            & ~~~~$69\pm35$   \\
Gonz (V$-$K)$^2$           & ~\,$-34\pm34$   \\
Casa (V$-$K)$^3$           & ~\,$-33\pm32$  \\
IRFM                 & $-108\pm47$ \\
\hline
Average $\sigma$        & ~~~~~~~\,$\pm$\,48           \\
\hline
\multicolumn{2}{l}{\footnotesize{$^1$\protect\shortcite{ramirez2005}}} \\
\multicolumn{2}{l}{\footnotesize{$^2$\protect\citet{gonzalez2009}}} \\
\multicolumn{2}{l}{\footnotesize{$^3$\protect\citet{casagrande2010}}}  \\
\end{tabular}
\end{table}

\subsection{Chemical abundance determination}
\label{6397_abund_method}

Chemical abundances were determined for Fe (using Fe\,{\sc i} and Fe\,{\sc ii}), Na (Na\,{\sc i}), O (O\,{\sc i}), Mg (Mg\,{\sc i}), and Al (Al\,{\sc i}) using the equivalent width (EW) method. EWs of absorption lines were measured using a combination of the {\sc ares} \citep[v2]{ares} and {\sc iraf} {\it onedspec} packages, while one-dimensional LTE abundances were determined using the {\sc moog} code \citep[June 2014 release]{moog} and model atmospheres that were interpolated from the \citet{atlas9odfnew} grid. The line list and atomic data used for this analysis are specified in Table~\ref{tab:6397_linelist}. The LTE assumption has been known for many years to be an inaccurate approximation for the abundances of many elemental species. In fact, all elements determined in this work are affected by non-LTE effects which must be accounted for if the abundances are to be reliable. Fortunately, grids of non-LTE corrections now exist for all of these elements in the parameter space occupied by our stellar sample.

Iron abundances determined from neutral absorption lines are known to be systematically lower than those determined using singly-ionised lines \citep[for which LTE is a realistic approximation;][]{lind2012}. However, due to the large number of Fe lines in a stellar spectrum, it can prove difficult to perform a complete line-by-line non-LTE analysis using published grids. For this reason, we performed a test to gauge the magnitude of the offsets on a subset of stars and lines. For our test, we selected a representative sub-sample of three RGB and three AGB stars from NGC\,6397, and interpolated corrections from \citet{amarsi2016_iron} for five Fe\,{\sc i} lines\footnote{4788.8\AA, 4839.5\AA, 5701.6\AA, 5753.1\AA~and 7748.3\AA} and two Fe\,{\sc ii} lines\footnote{6516.1\AA~and 7711.7\AA}. The  results of this test are summarised the first two rows of Table~\ref{tab:6397_nlte}. We did not apply these average corrections, but compare them to our LTE Fe results in Section~\ref{6397_results}. 

Non-LTE corrections were applied to all Na, O, Mg and Al abundances line-by-line using the most recent grids. As in ML16, Na abundances were determined using the 568~nm doublet and corrected for non-LTE effects as described in \citet{lind2011} by using the web-based {\sc inspect} interface\footnote{http://inspect-stars.net}, and adopting the provided $\Delta$[Na/Fe]$_{\text{\sc nlte}}$ corrections. The oxygen 777~nm triplet was measured and non-LTE corrections were determined by the interpolation of the recent \citet{amarsi2016_oxygen} grid of corrections. For Mg, the measured EWs of the 571~nm and 769~nm lines were used for non-LTE determinations as described in \citet{osorio2016}, using the {\sc inspect} interface. The average of these two values was then used to correct the 473~nm Mg line. Finally, both the 669~nm and 783~nm doublets were used to determine Al abundances, while non-LTE adjustments were interpolated from the new results of \citet{nordlander2017}. Average non-LTE corrections, and associated spreads are listed in Table~\ref{tab:6397_nlte}.

\begin{table}
\centering
\caption{Adopted line list used for EW measurements. Based on the line list of the GALAH collaboration \protect{\citep{galah}}.}
\label{tab:6397_linelist}
\begin{tabular}{cccr}
\hline
Wavelength & Species & Excitation Potential & log $gf$ \\
(\AA) &  & (eV) &  \\
\hline\hline
7771.94 & O\,{\sc i}    & 9.146 & ~0.369  \\
7774.16 & O\,{\sc i}    & 9.146 & ~0.223  \\
7775.39 & O\,{\sc i}    & 9.146 & ~0.002  \\
5682.63 & Na\,{\sc i}   & 2.100   & $-0.706$ \\
5688.20 & Na\,{\sc i}   & 2.100   & $-0.404$ \\
4730.03 & Mg\,{\sc i}   & 4.350  & $-2.347$ \\
5711.09 & Mg\,{\sc i}   & 4.350  & $-1.724$ \\
7691.53 & Mg\,{\sc i}   & 5.750  & $-0.783$ \\
6696.02 & Al\,{\sc i}   & 3.140  & $-1.569$ \\
6698.67 & Al\,{\sc i}   & 3.140  & $-1.870 $ \\
7835.31 & Al\,{\sc i}   & 4.020  & $-0.689$ \\
7836.13 & Al\,{\sc i}   & 4.020  & $-0.534$ \\
4788.76 & Fe\,{\sc i}   & 3.237 & $-1.763$ \\
4839.54	& Fe\,{\sc i} 	& 3.270	& $-1.820$ \\
4890.75 & Fe\,{\sc i}   & 2.875 & $-0.394$ \\
4891.49 & Fe\,{\sc i}   & 2.849 & $-0.111$ \\
5701.56 & Fe\,{\sc i}   & 2.559 & $-2.220 $ \\
5753.12	& Fe\,{\sc i}	& 4.260	& $-0.690$ \\
5859.59 & Fe\,{\sc i}   & 4.549 & $-0.419$ \\
5862.36 & Fe\,{\sc i}   & 4.549 & $-0.127$ \\
6498.94 & Fe\,{\sc i}   & 0.958 & $-4.687$ \\
6518.37 & Fe\,{\sc i}   & 2.831 & $-2.440$  \\
6592.91 & Fe\,{\sc i}   & 2.727 & $-1.473$ \\
6593.87 & Fe\,{\sc i}   & 2.433 & $-2.420$  \\
6609.11 & Fe\,{\sc i}   & 2.559 & $-2.691$ \\
6677.99 & Fe\,{\sc i}   & 2.690  & $-1.420$  \\
7748.27 & Fe\,{\sc i}   & 2.949 & $-1.751$ \\
7780.56 & Fe\,{\sc i}   & 4.473 & $-0.010$  \\
4731.45 & Fe\,{\sc ii} & 2.891 & $-3.100$   \\
6516.08 & Fe\,{\sc ii} & 2.891 & $-3.310$  \\
7711.72 & Fe\,{\sc ii} & 3.903 & $-2.500$   \\
\hline
\end{tabular}
\end{table}

\begin{table}
\centering
\caption{Summary of average non-LTE corrections for each element, with 1$\sigma$ standard deviations over the stellar sample.}
\label{tab:6397_nlte}
\begin{tabular}{ccc}
\hline
Species & \multicolumn{2}{c}{Average non-LTE Correction} \\
 & RGB & AGB \\
\hline\hline
Fe\,{\sc i} & $+0.08$ $\pm$ 0.04 & $+0.08$ $\pm$ 0.03 \\
Fe\,{\sc ii} & $<0.01$ & $<0.01$ \\
O  & $-0.05$ $\pm$ 0.01 & $-0.06$ $\pm$ 0.01 \\
Na & $-0.06$ $\pm$ 0.02 & $-0.06$ $\pm$ 0.01 \\
Mg & $+0.02$ $\pm$ 0.01 & $+0.02$ $\pm$ 0.01 \\
Al & $-0.06$ $\pm$ 0.03 & $-0.05$ $\pm$ 0.05 \\
\hline
\end{tabular}
\end{table}


\section{Abundance results \& analysis}
\label{6397_results}

Final elemental abundances are presented in Table~\ref{tab:6397_abunds}. Uncertainties cited in the table are based only on the line-to-line scatter of each abundance and do not consider additional sources of error. Using our estimated 1$\sigma$ uncertainties of each stellar parameter ($\pm$50K in $T_{\rm eff}$, $\pm$0.1 in log $g$, $\pm$0.2 km/s in $v_{\rm t}$), an atmospheric sensitivity analysis was performed on a representative sub-sample and results are summarised in Table~\ref{tab:6397_atmos}. Finally, in Table~\ref{tab:6397_uncerts} we present a summary of all identified sources of uncertainties and adopted total abundance uncertainties.

\begin{table*}
\centering
\caption{Stellar parameters, and derived chemical abundances for each star in NGC\,6397. Abundance uncertainties reflect line-to-line scatter (1$\sigma$), and do not take atmospheric sensitivities into account (see Table~\ref{tab:6397_atmos}, and text for discussion). The last two rows are the cluster average abundances with error on the mean, and standard deviation to indicate observed scatter. We adopt the \protect{\citet{asplund2009}} solar abundance values. The full table is available online.}
\label{tab:6397_abunds}
\begin{tabular}{ccccccccccc}
\hline
ID  & Type & $T_{\rm eff}$ & log $g$ & $v_{\rm t}$   & {[}Fe\,{\sc i}/H{]}     & {[}Fe\,{\sc ii}/H{]}    & {[}O/H{]}        & {[}Na/H{]}       & {[}Mg/H{]}       & {[}Al/H{]}       \\
& & (K) & (cgs) & (km/s) & & & &  \\
\hline\hline
56897 & AGB  & 4978 & 1.80 & 1.64 & $-2.13 \pm 0.06$ & $-2.00 \pm 0.02$ & $-1.64 \pm 0.01$ & $-1.92 \pm 0.01$ & $-1.84 \pm 0.04$ & $-1.32 \pm 0.03$ \\
60609 & AGB  & 4905 & 1.70 & 1.67 & $-2.23 \pm 0.07$ & $-2.06 \pm 0.01$ & $-1.45 \pm 0.04$ & $-1.98 \pm 0.01$ & $-2.02 \pm 0.01$ & $-1.37 \pm 0.01$ \\
70509 & AGB  & 5017 & 1.88 & 1.61 & $-2.18 \pm 0.06$ & $-2.07 \pm 0.04$ & $-1.49 \pm 0.04$ & $-2.15 \pm 0.01$ & $-1.79 \pm 0.05$ & $-1.53 \pm 0.04$ \\
70522 & AGB  & 4739 & 1.42 & 1.76 & $-2.24 \pm 0.05$ & $-2.06 \pm 0.03$ & $-1.63 \pm 0.02$ & $-1.94 \pm 0.04$ & $-1.99 \pm 0.08$ & $-1.48 \pm 0.06$ \\
73216 & AGB  & 4968 & 1.80 & 1.64 & $-2.16 \pm 0.05$ & $-2.04 \pm 0.00$ & $-1.39 \pm 0.06$ & $-2.29 \pm 0.04$ & $-1.73 \pm 0.05$ & $-1.67 \pm 0.03$ \\
\vdots & \vdots & \vdots & \vdots & \vdots & \vdots & \vdots & \vdots & \vdots & \vdots & \vdots  \\
\hline
Mean & & & & & $-2.15 \pm 0.01$ & $-2.02 \pm 0.00$ & $-1.52 \pm 0.02$ & $-2.06 \pm 0.02$ & $-1.87 \pm 0.01$ & $-1.49 \pm 0.02$ \\
$\sigma$ & & & & & 0.05 & 0.03 & 0.12 & 0.19 & 0.11 & 0.16 \\
\hline
\end{tabular}
\end{table*}

\begin{table}
\centering
\caption{Typical abundance uncertainties due to the (1$\sigma$) atmospheric sensitivities of a representative sub-sample of three RGB and two AGB stars in our NGC\,6397 data set. Parameter variations (in parentheses) are the expected uncertainties in the respective parameters.}
\label{tab:6397_atmos}
\begin{tabular}{lccccc}
\hline
 & ${\Delta}T_{\rm eff}$          & ${\Delta}$log $g$    & ${\Delta}v_{\rm t}$         & \bf Total       \\
 & ($\pm$50 K) & ($\pm$0.1 dex) & ($\pm$0.2 km/s) \\
\hline\hline
{[}Fe\,{\sc i}/H{]}  & $\pm$0.06  & $\mp$0.01  & $\mp$0.04  & \bf $\pm$0.04 \\
{[}Fe\,{\sc ii}/H{]} & $\mp$0.01  & $\pm$0.04  & $\mp$0.01  & \bf $\pm$0.04 \\
{[}O/H{]}     		& $\mp$0.06   & $\pm$0.04  & $\pm$0.00  & \bf $\pm$0.07 \\
{[}Na/H{]}    		& $\pm$0.03   & $\mp$0.01  & $\pm$0.00  & \bf $\pm$0.03 \\
{[}Mg/H{]}   		& $\pm$0.03   & $\mp$0.00  & $\pm$0.00  & \bf $\pm$0.03 \\
{[}Al/H{]}    		& $\pm$0.03   & $\pm$0.00  & $\pm$0.00  & \bf $\pm$0.03 \\
\hline
\end{tabular}
\end{table}

\begin{table}
\centering
\caption{Summary of typical abundance uncertainties (1$\sigma$) from each source identified in the text, and the total uncertainties (added in quadrature). The first column are the average line-to-line uncertainties of all stars, values in the second column are the total uncertainties from atmospheric sensitivities (Table~\ref{tab:6397_atmos}), and the third column represents the typical uncertainties quoted in each non-LTE source (see {\S}\ref{6397_abund_method} for citations). Note that individual Fe abundances were not corrected for non-LTE (see text for details).}
\label{tab:6397_uncerts}
\begin{tabular}{ccccc}
\hline
Species & Line-to-Line & Atmospheric & non-LTE & \bf Total \\
\hline\hline
Fe\,{\sc i} & $\pm$0.07 & $\pm$0.04 & - & \bf $\pm$0.08 \\
Fe\,{\sc ii} & $\pm$0.03 & $\pm$0.04 & - & \bf $\pm$0.05 \\
O  & $\pm$0.03 & $\pm$0.07 & $\pm$0.05 & \bf $\pm$0.09 \\
Na & $\pm$0.04 & $\pm$0.03 & $\pm$0.04 & \bf $\pm$0.06 \\
Mg & $\pm$0.04 & $\pm$0.03 & $\pm$0.03 & \bf $\pm$0.06 \\
Al & $\pm$0.04 & $\pm$0.03 & $\pm$0.06 & \bf $\pm$0.08 \\
\hline
\end{tabular}
\end{table}

A comparison of our results was made with that of \citet[L11]{lind2011} and \citet[C09]{carretta2009vii}, with which we had a total of 5 and 21 RGB stars in common, respectively. The results of the detailed comparison of all stellar parameters and abundances are presented in Table~\ref{tab:6397_overlap}, which shows good agreement in all stellar parameters and slight to moderate offsets in abundance results (0.03 to 0.18 dex) between the studies. These offsets arise from different methods in analysis.

In the cases of assumed stellar mass, atmospheric model parameters, adopted non-LTE corrections, and adopted solar abundances, we were able to quantify the effects since the previous studies published their values for these inputs. These sources of uncertainty combine to total possible offsets of up to $+0.10$ dex in each abundance. Other sources of uncertainty  which we could not quantify (because we do not have the relevant information from the related studies) -- for example different line lists, EW measurements and instrumentation differences -- most likely explain the remaining offsets. We note that the scatter around these offsets is typically considered a better indication of the agreement between abundance analysis studies, and is consistent with the uncertainties quoted in this work. We find very good agreement between our study and that of L11. A curiosity here is the lack of agreement on micro-turbulence values with C09. While we adopted photometric $v_{\rm t}$ (and therefore had a relatively small spread in values, ranging from 1.52 km/s to 1.71 km/s), C09 determined micro-turbulence spectroscopically and had a very large spread in $v_{\rm t}$ values (ranging from 0.11 km/s to 2.73 km/s in the overlapping sample). This may explain the increased offsets and scatter between C09 and our study.

\begin{table}
\centering
\caption{The average differences in parameters and abundances between this work and that of \protect{\citet[L11]{lind20116397}} and \protect{\citet[C09]{carretta2009vii}}. Uncertainties are standard deviations, and indicate the scatter around the offsets. While significant offsets exist between our work and the works of L11 and C09, the scatter around the offsets are consistent with the uncertainties quoted in this work (see text for discussion).}
\label{tab:6397_overlap}
\begin{tabular}{lcc}
\hline
Parameter & This study $-$ L11   & This study $-$ C09               \\
\hline\hline
$\Delta T_{\rm eff}$     		& ~\,$-4.3$ $\pm$ 20.9    & ~~\,19.5 $\pm$ 29.9        \\
$\Delta$log $g$ 			& ~~\,0.08 $\pm$ 0.01 & ~~\,0.07 $\pm$ 0.02 	 \\
$\Delta v_{\rm t}$ 			& ~~\,0.04 $\pm$ 0.03  & ~~\,0.21 $\pm$ 0.64 	\\
\hline
$\Delta${[}Fe\,{\sc i}/H{]}  & $-0.08$ $\pm$ 0.03  & $-0.13$ $\pm$ 0.05 	\\
$\Delta${[}Fe\,{\sc ii}/H{]}  & ~~\,0.12 $\pm$ 0.03  & ~~\,0.05 $\pm$ 0.05		\\
$\Delta${[}O/H{]}  		& ~~\,0.06 $\pm$ 0.08  & ~~\,0.18 $\pm$ 0.14  		\\
$\Delta${[}Na/H{]} 		& $-0.03$ $\pm$ 0.06 & $-0.17$ $\pm$ 0.14 		\\
$\Delta${[}Mg/H{]} 		& $-0.09$ $\pm$ 0.03 & ~~\,- 		\\
$\Delta${[}Al/H{]} 		& ~~\,0.17 $\pm$ 0.12 & ~~\,- 		\\
\hline
\end{tabular}
\end{table}

\begin{figure}
\centering
\includegraphics[width=0.9\linewidth]{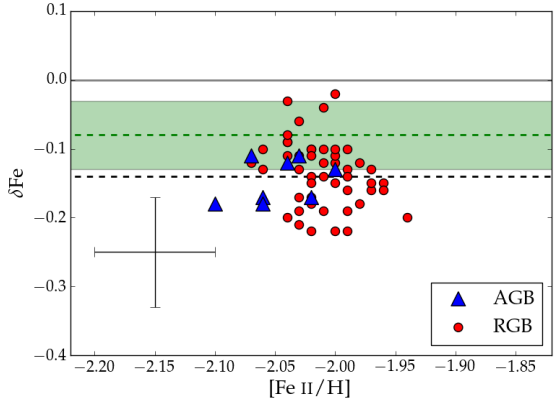}
\caption{LTE Fe abundances for our NGC\,6397 sample. Here, $\delta$Fe ([Fe\,{\sc i}/H] $-$ [Fe\,{\sc ii}/H]) is plotted against [Fe\,{\sc ii}/H] abundance to highlight departures from LTE in Fe\,{\sc i}, and the similarity between the Fe abundances of the AGB and RGB. The error bars indicate typical 1$\sigma$ total uncertainties on individual abundances (see Table~\ref{tab:6397_uncerts}), while the black dashed line represents the sample average $\delta$Fe value of $-0.14$ dex. The green dashed line represents the expected $\delta$Fe value ($-0.08$ dex) from our non-LTE test (see \S\ref{6397_abund_method}) and the shaded region indicates the non-LTE uncertainties quoted in \protect{\citet[$\pm$0.05 dex]{amarsi2016_iron}}.}
\label{fig:6397_dfe}
\end{figure}

The difference between our mean LTE {[}Fe\,{\sc i}/H{]} and {[}Fe\,{\sc ii}/H{]} abundances\footnote{$\delta$Fe~$=$~[Fe\,{\sc i}/H] $-$ [Fe\,{\sc ii}/H]} (from Table~\ref{tab:6397_abunds}) across our sample is {\textless}$\delta$Fe{\textgreater}$ = -0.14 \pm 0.01$ (${\sigma}=0.05$). This is 0.06 dex lower lower than value predicted by non-LTE theory ($-0.08 \pm 0.05$ dex, see \S\ref{6397_abund_method} and Fig~\ref{fig:6397_dfe}). While this could indicate slight systematics in either our $T_{\rm eff}$ estimates or the non-LTE corrections, the uncertainty range of our {\textless}$\delta$Fe{\textgreater} value overlaps with that of the the non-LTE predicted $\delta$Fe value, indicating broad agreement. Our Fe abundances are consistent with literature values ({\textless}[Fe/H]{\textgreater}$_{\rm L11}$ = $-2.08 \pm 0.02$). Furthermore, the difference between the average RGB and AGB $\delta$Fe values is less than 0.015 dex for NGC\,6397, indicating that there are no significant offsets in $\delta$Fe between the two giant branches, as has been disputed for NGC\,6752 \citep{lapenna2016,campbell2017}. This is presented visually in Figure~\ref{fig:6397_dfe}, where the overall homogeneity of Fe abundances can be seen, especially between the AGB and RGB.

\begin{figure}
\centering
\includegraphics[width=0.9\linewidth]{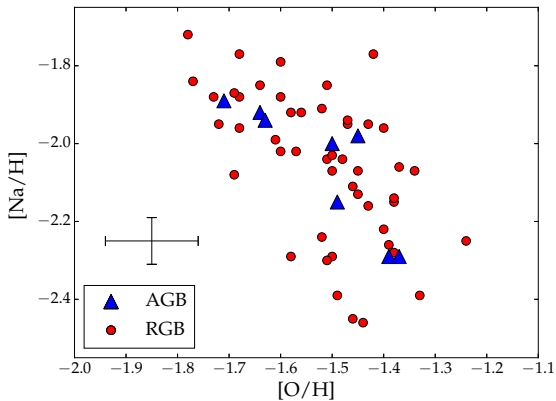}
\caption{Na and O abundances for our NGC\,6397 sample. The error bars indicate typical 1$\sigma$ total uncertainties on individual abundances (see Table~\ref{tab:6397_uncerts}).}
\label{fig:6397_nao}
\end{figure}

\begin{figure}
\centering
\includegraphics[width=0.9\linewidth]{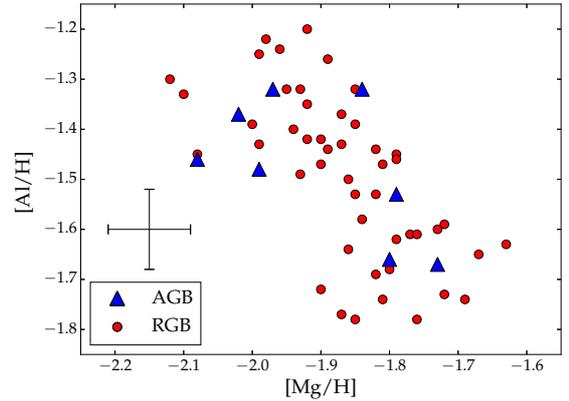}
\caption{Same as Figure~\ref{fig:6397_nao}, but for Mg and Al.}
\label{fig:6397_mgal}
\end{figure}

\begin{figure}
\centering
\includegraphics[width=0.9\linewidth]{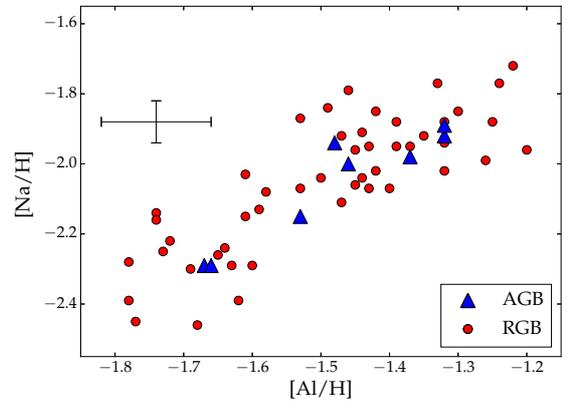}
\caption{Same as Figure~\ref{fig:6397_nao}, but for Na and Al.}
\label{fig:6397_naal}
\end{figure}

Abundances of elements other than iron are presented in Figures~\ref{fig:6397_nao},~\ref{fig:6397_mgal},~and~\ref{fig:6397_naal}. NGC\,6397 was shown by L11 to have both Na-O and Mg-Al anti-correlations, which we find on both the RGB and AGB, along with a Na-Al correlation (Fig~\ref{fig:6397_naal}). The abundance distributions of the two giant branches in NGC\,6397 are remarkably similar -- we find that $\mathscr{R}_{RGB} \simeq \mathscr{R}_{AGB} \simeq 60\%$ (compared with $\mathscr{R}_{RGB} \simeq 75\%$ in L11), indicating no SP2 AGB deficit, i.e.; $\mathscr{F} \simeq 0\%$. This is in agreement with current stellar evolutionary theory, which predicts that there should be no AGB-manqu\'{e} stars in NGC\,6397, due to a HB that only extends to $T_{\rm eff} \simeq 10,500$~K \citep{agb-manque,dorman1993,lovisi2012}.

\begin{figure}
\centering
\includegraphics[width=0.9\linewidth]{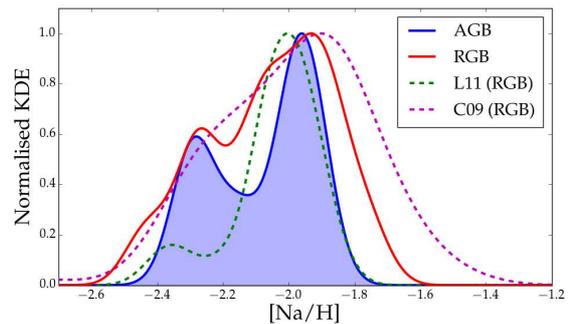}
\caption{Gaussian kernel density estimations (KDEs) of our NGC\,6397 [Na/H] abundances, along with those of \protect{\citet[L11]{lind20116397}} and \protect{\citet[C09]{carretta2009vii}}, with systematic offsets removed (see text for details). A smoothing bandwidth of 0.06 dex (total Na uncertainty, see Table~\ref{tab:6397_uncerts}) was applied to each of our RGB and AGB data sets, while for C09 we used a bandwidth of 0.11 dex, matching their total error calculations (see C09, Appendix A). L11 did not quote total abundance uncertainties, however their average measurement uncertainty in Na was the same as in our sample (0.04 dex), therefore we applied an identical bandwidth of 0.06 dex. The discrepancy between the relative heights of the two peaks in the L11 sample, compared to those of the other samples, may be due to the low number of stars observed in L11 (21 RGB stars).}
\label{fig:6397_kde}
\end{figure}

Finally, in Figure~\ref{fig:6397_kde}, we present Gaussian kernel density estimations (KDEs) of our AGB and RGB samples. We also plot KDEs of the RGB samples from L11 and C09 for comparison. Constant corrections of $-0.03$ dex and $-0.17$ dex, respectively, were applied to the data of these studies based on the systematic [Na/H] offsets determined (Table~\ref{tab:6397_overlap}). Figure~\ref{fig:6397_kde} shows excellent agreement between the [Na/H] abundances of our current RGB and AGB samples, as well as between our RGB results and the RGB results of L11 and C09.

\section{Discussion and conclusions}
\label{6397_discussion}

The primary goal of this study was to determine the proportion of SP2 stars in NGC\,6397 that evolve through to the AGB phase. Since the work of \citet{campbell2013}, the nature of AGB stars in GCs has been debated in the literature, with eight high-resolution spectroscopic studies \citep{johnson2015,garciahernandez2015,lapenna2015,maclean2016,lapenna2016,wang2016,marino2017,wang2017} and five photometric studies \citep{monelli2013,milone2015_7089,milone2015_2808,lardo2017,gruyters2017} targeting the AGB directly, along with five theoretical studies seeking to explain the anomalous observations \citep{charbonnel2013,cassisi2014,charbonnel2014,charbonnel2015,charbonnel2016}.

Since only HB stars with effective temperatures above $\sim$15,000~K are predicted to evolve directly to the white dwarf phase, the AGBs of clusters that lack an extended blue HB are expected to contain distributions in Na, O, Mg and Al abundances that are statistically indistinguishable from those of the RGB -- all cluster stars should evolve through both giant branches (i.e., $\mathscr{F} \simeq 0\%$). Only in clusters with extended blue HBs should the distribution be different, and only with the $\sim$30 per cent most extreme (Na-rich/O-poor/Al-rich) AGB stars missing \citep[i.e., $\mathscr{F} \simeq 30\%$][]{dorman1993,cassisi2014}.

Despite a rapidly expanding literature sample of GC AGB studies, the picture is still far from clear. To date, eleven GCs have had their AGB systematically probed with high-resolution spectrographs\footnote{NGC\,2808, NGC\,6397, NGC\,6752, 47 Tucanae, M\,2, M\,3, M\,4, M\,5, M\,13, M\,55 \& M\,62}, with mixed results in $\mathscr{F}$ values (see ML16, Table~4). However, only three clusters have been reported to have $\mathscr{F} \simeq 100\%$: NGC\,6752 \citep[][C17]{campbell2013}, M\,62 \citep{lapenna2015} and M\,4 (ML16). Of these, only M\,62 has not been disputed by subsequent studies, but we note that this GC has not yet been studied a second time.

\citet{lapenna2016} reported that the Fe\,{\sc i} abundances of AGB stars in NGC\,6752 are lower than predicted by standard non-LTE theory. If extrapolated to Na abundance, (i.e., if Na is assumed to follow this trend), the AGB [Na{\sc i}/Fe{\sc i}] abundance distribution moves to be in line with stellar theory ($\mathscr{F} \simeq 30\%$, as expected in GCs with an extended blue HB), contradicting the conclusions of \citet{campbell2013} who claimed $\mathscr{F} \simeq 100\%$. However, in a detailed re-analysis of their data, C17 reported that there was no iron abundance discrepancy in NGC\,6752 when more reliable $T_{\rm eff}$ scales were used, therefore concluding that the original Na results of \citet{campbell2013} are reliable. Furthermore, for NGC\,6397 we have found no significant $\delta$Fe offset between the AGB and RGB, and that the Fe abundances are internally homogeneous (at the level of our uncertainties). This allows [X/H] abundances to be used for the elemental distribution analyses of the giant branches, because using [X/Fe] would introduce additional scatter (through measurement uncertainties), but no new information.

The abundances of NGC\,6397 (Figs~\ref{fig:6397_nao}-\ref{fig:6397_kde}) contain no evidence of a SP2 AGB deficit, with the relative distributions of the RGB and AGB being identical in all abundance planes ($\mathscr{F} \simeq 0\%$). 

It is interesting to compare this result with that of M\,4 by ML16, since the methods and tools we have used are almost identical. The only difference between the NGC\,6397 analysis performed in this study and that of ML16 is the method of determining atmospheric parameters. In ML16, $T_{\rm eff}$, log $g$ and $v_{\rm t}$ values were determined spectroscopically by requiring excitation and ionisation balance \citep[as per][]{sousa2014}, whereas for NGC\,6397 these parameters were estimated through photometric relations. As shown in C17, Na{\sc i} abundances are quite robust, that is they are not as sensitive to systematic shifts in $T_{\rm eff}$ as Fe{\sc i} abundances.  We have also shown that our Fe results are consistent with non-LTE theory, and show homogeneous abundances in both ionisation states, indicating that our $T_{\rm eff}$ scale is accurate. For these reasons, we consider that the different method of parameter determination between our two studies should have little consequence on the reliability of our [Na/H] abundances.

Thus our NGC\,6397 result further strengthens the conclusions of ML16 whose analysis was almost identical, but whose results are in contradistinction. We therefore suggest that our original M\,4 conclusions ($\mathscr{F} \simeq 100$, but with some uncertainty) are sound, and that our NGC\,6397 results show -- by providing a control sample -- that our method of analysis does not artificially shift AGB abundances toward SP1-like distributions.

As stated in ML16, our M\,4 result ($\mathscr{F} \simeq 100$) is in clear contradiction with stellar theory -- we can think of no reason why SP2 stars in M\,4 should avoid the AGB phase, since the maximum $T_{\rm eff}$ of its HB is $\sim$9000 K \citep{marino2011}. This is especially true in light of our result for NGC\,6397 -- which has a bluer HB than M\,4, but $\mathscr{F} \simeq 0\%$. In the search for a possible explanation of our results, and those of \citet[NGC\,6752]{campbell2013,campbell2017} and \citet[M\,62]{lapenna2015}, we consider three possible causes of the low-Na signature of AGB stars in M\,4, NGC\,6752 \& M\,62:
\renewcommand\labelenumi{(\roman{enumi})}
\renewcommand\theenumi\labelenumi
\begin{enumerate}
\item The low-Na signature is intrinsic -- HB stars are becoming AGB-manqu\'{e} stars at a much lower HB $T_{\rm eff}$ than predicted. This is the most commonly cited explanation in the literature.
\item The atmospheric models of some AGB stars are incorrectly determined, but only in particular sections of the GC AGB parameter space. This would result in incorrectly predicted absorption line profiles, and represent a significant `blind spot' in the standard spectroscopic method.
\item All Na-rich stars in these three GCs are undergoing an unknown burning or mixing process, between the HB and AGB, that acts to deplete Na in the envelope and leave only a low-Na signature by the early AGB phase.
\end{enumerate}

Investigating these hypotheses is beyond the scope of the present work. However, we note that (iii) is almost certainly impossible since there is no known mechanism that can destroy Na, while simultaneously creating O, in the interior conditions found in these stars.

More generally we note that, of the GCs which have been analysed for SP2 AGB deficits, not a single deficit (or lack thereof) claim has been confirmed by a different working group, or with independently selected targets. This suggests that the methods that are used require detailed investigation and checking, such as performed in C17. This is especially pertinent for M\,4, for which the three existing studies all give different values of $\mathscr{F}$. We will aim to resolve this issue in a forthcoming study. Finally, we suggest another potential next step in investigating this problem could be a controlled spectroscopic study of an `HB second parameter' pair or trio of clusters with similar metallicity and age, but different HB-morphology (such as NGC\,288, NGC\,362, and NGC\,1851), in an attempt to disentangle the effect of global GC parameters on apparent AGB deficits.


\section*{Acknowledgements}

Based in part on data acquired through the AAO, via program 15A/21 (PI Campbell). Part of this work was supported by the DAAD (PPP project 57219117) with funds from the German Federal Ministry of Education and Research (BMBF). BTM acknowledges the financial support of his Australian Postgraduate Award scholarship. SWC acknowledges federal funding from the Australian Research Council though the Future Fellowship grant entitled ``Where are the Convective Boundaries in Stars?'' (FT160100046). VD acknowledges support from the AAO distinguished visitor program 2016. LC gratefully acknowledges support from the Australian Research Council (grants DP150100250, FT160100402). We thank Anish Amarsi and Thomas Nordlander for providing non-LTE corrections, and Michael Brown, Alexander Heger, and the referee for useful conversations, comments, and advice.

\bibliography{References}


\label{lastpage}

\end{document}